\documentclass[prb,onecolumn,notitlepage,eqsecnum,nofootinbib,floatfix]{revtex4-1}

\usepackage{amsmath}  
\usepackage{amsfonts} 
\usepackage{graphicx} 

\usepackage{simplewick}

\usepackage{subcaption}

\begin{document}


\title{Six textbook mistakes in quantum field theory}

\author{Alexandros Gezerlis}
\affiliation{Department of Physics, University of Guelph, Guelph, Ontario N1G 2W1, Canada}

\date{\today}

\begin{abstract}
This article discusses incorrect statements appearing in textbooks on quantum field theory (QFT); some of these mistakes also appear in the research literature. 
The focus is not on errors made by an individual author, but on conceptual muddledness that is widespread in introductory textbooks.  
We start from a bare-bones summary of QFT, meant to establish the notation.
We then turn to our six paradigmatic themes, in each case quoting a specific example
of the textbook mistake, a summary 
of material that is known to experts but is frequently mishandled in
introductory works, pointers to authoritative references where the relevant concept is handled properly, as well as a concise correction that rectifies any issues. 
The goal of this work 
is to warn readers of the existence of several pitfalls and thereby
stop these errors from further propagating in the literature on QFT.
\end{abstract}

\maketitle 

\section{Introduction}

This is the third installment in a series of works\cite{GezerlisWilliams1,GezerlisWilliams2} tackling mistaken claims appearing in physics textbooks. The premise of these articles is that errors in the research literature are both inevitable and
understandable, but textbooks should be held to a higher standard. Not only because their writing has the benefit of hindsight (unlike journal publications, which typically deal with questions that are rapidly evolving) but also, 
and even more importantly, because textbooks reach a much broader (and much more impressionable) audience.
The present article studies widespread misconceptions on themes related to quantum field theory, which we also take to include the usual preliminaries (relativistic quantum mechanics and classical field theory). Relativistic quantum field theory is typically taught in a dedicated graduate course, often over two semesters; 
the misconceptions we discuss revolve around ``core'' aspects of QFT, so they
are mainly of relevance to the first semester (or even an undergraduate version)
of such a course.

In order to structure this work, we have employed a selection criterion: 
each mistake addressed appears in at least two standard textbooks.
Another criterion is that the questions touched upon here are of wide conceptual 
import, not one-off miscalculations.
Sometimes individual textbook authors make mistakes (even glaring ones), but that is not our focus here: we care about incorrect claims that are widespread, i.e.,
ones which have been propagating through the introductory literature (despite the fact
that experts would ordinarily not make such claims).
Typos or other minor issues can typically be addressed by a student on the fly, 
while working through a textbook. On the other hand, conceptual misunderstanding, especially when it is not limited to a single textbook, is much more pernicious, often leading students to blame their own thought process for their inability to understand what is going on. This is even more true for a forbidding subject like quantum field theory, which already has a reputation of being expert-friendly.

The previous two installments in this series grew out of the author's work on 
writing\cite{GezerlisNumerical1} (or updating)\cite{GezerlisNumerical2} a physics textbook, a process which was preceded by an in-depth exploration of the introductory textbook literature. The provenance of the present article is similar, 
in that the author has recently finished writing a textbook on quantum field
theory.\cite{GezerlisQFT}
The six topics addressed in this work are tackled correctly in 
Ref.\ \onlinecite{GezerlisQFT} (in more detail than here), 
but readers can always benefit from having
multiple reliable sources on a given subject, so in what follows 
we will also cite authoritative works by other authors on each theme.
The motivation behind the present article is that focusing on these widespread
incorrect claims will, it is to be hoped, make it possible for future generations of students 
to learn the subject right the first time. 
Our intended audience is mostly composed of QFT instructors, who 
may have unintentionally contributed toward the propagation of
these textbook mistakes in the past.

\section{Establishing the notation}
\label{sec:notation}

Many of the textbook mistakes to be discussed below 
arise because the notation used in quantum field theory is often too sloppy (albeit sometimes for good reason). 
For example, it is very common to denote a real classical field by $\phi$, 
but then also use the same symbol for a complex classical field, 
or a quantum field in the Heisenberg picture (or in another picture). 
In order to preempt any misunderstanding about such matters, 
here we first go over the notation that we will be employing in the
rest of the article. We emphasize that this section is meant as reference (not pedagogical) material; trying to teach quantum field theory from scratch 
in a few pages would be a doomed exercise.

As is standard in QFT, we will be working with natural units, setting $\hbar = c = 1$. 
We will be using the particle physics/mostly minus metric $\eta_{\mu \nu}$,
with signature $(+, -, -, -)$, 
according to which
$a^{\mu} b_{\mu} = a^0 b^0 - \mathbf{a} \cdot \mathbf{b}$. 
The Lagrangian $L$ is related to the Lagrangian density $\cal{L}$ via 
$L = \int d^3x \cal{L}$. In what follows, we take ${\cal L} = {\cal L}(\phi, \partial_{\mu} \phi)$, for a real classical field $\phi = \phi(t,\mathbf{x}) = \phi(x)$. By employing an intrinsic variation:
\begin{equation}
\phi'(x) = \phi(x) + \delta\phi(x)
\label{eq:intrinsic_first}
\end{equation}
we can apply Hamilton's principle to find the Euler--Lagrange equation:
\begin{equation}
\partial_{\mu} \frac{\partial {\cal L}}{\partial 
(\partial_{\mu} \phi)} - \frac{\partial {\cal L}}{\partial \phi} = 0
\label{eq:scalar_Lag}
\end{equation}
If we now plug the Lagrangian density:
\begin{equation}
{\cal L} = \frac{1}{2} (\partial \phi)^2  - \frac{1}{2} m^2 \phi^2
\label{eq:scalar_II}
\end{equation}
where $(\partial \phi)^2 \equiv \partial_{\mu} \phi \partial^{\mu} \phi$, into the Euler--Lagrange equation, we find the 
following field equation:
\begin{equation}
\left ( \partial^2 + m^2 \right ) \phi = 0
\label{eq:KG_ch2}
\end{equation}
which is known as the Klein--Gordon equation. Its solutions are plane waves
with energy dispersion $E^2 = \mathbf{p}^2 + m^2$ or, more generally,
the field Fourier-mode expansion:
\begin{equation}
\phi(t,\mathbf{x}) =  \int \frac{d^3k}{(2\pi)^3 2\omega} \left [ a(\mathbf{k}) e^{-ikx}
+ a^*(\mathbf{k}) e^{ikx} \right ]
\label{eq:scalar_field_exp}
\end{equation}
where we are employing a Lorentz-invariant integration measure and 
the angular frequency is $\omega = \sqrt{\mathbf{k}^2 + m^2}$.

Still at the level of classical field theory, 
we can examine what happens under a more
general transformation than that in Eq.~(\ref{eq:intrinsic_first}).
Specifically, we introduce the total variation:
\begin{equation}
\phi'(x') = \phi(x) + \tilde{\delta}\phi(x)
\label{eq:change_total}
\end{equation}
where 
\begin{equation}
{x'}^{\mu} = x^{\mu} + \delta x^{\mu}
\label{eq:change_coor}
\end{equation}
A transformation wherein we go from ${\cal L}(\phi(x), \partial \phi(x)/\partial x^{\mu})$
to ${\cal L}(\phi'(x'), \partial \phi'(x')/\partial x'^{\mu})$ 
while at the same time leaving the action functional
(${\cal S} = \int dt L$) invariant is known as a symmetry. 
A crucial result in this connection is Noether’s theorem: every continuous global symmetry transformation leads to a conserved current. Explicitly, we have:
\begin{equation}
j^{\mu} \equiv \left [ \frac{\partial {\cal L}}{\partial (\partial_{\mu} \phi)} \partial_{\nu} \phi - \eta_{\nu}^{\mu} {\cal L} \right ] \delta x^{\nu} - \frac{\partial {\cal L}}{\partial (\partial_{\mu} \phi)} \tilde{\delta} \phi
\label{eq:Noether_first_curr}
\end{equation}
and $\partial_{\mu} j^{\mu} = 0$.

Classically (still), we can pass to the 
Hamiltonian formalism, employing 
the Hamiltonian $H$ and the Hamiltonian density $\cal{H}$, related via 
$H = \int d^3x \cal{H}$. This is connected to the Lagrangian formalism
via the Legendre transform:
\begin{equation}
{\cal H}(x) \equiv \pi(x) \dot{\phi}(x) - {\cal L}(x) 
\label{eq:Ham_def_scalar}
\end{equation}
where the canonically conjugate momentum density $\pi(x)$ is given 
in terms of a functional derivative of the Lagrangian:
\begin{equation}
\pi(t,\mathbf{x}) \equiv \frac{\delta L}{\delta \dot{\phi}(t,\mathbf{x})}
\label{eq:scalar_mom_0}
\end{equation}
Crucially, if we apply Noether's theorem to the case of time translation,
the conserved charge is precisely the Hamiltonian $H$.
For the case of Eq.~(\ref{eq:scalar_II}), Eq.~(\ref{eq:scalar_mom_0}) leads to $\pi = \dot{\phi}$
and therefore:
\begin{align}
{\cal H} = \frac{1}{2} \pi^2(x) + \frac{1}{2} (\nabla \phi(x))^2  
+ \frac{1}{2} m^2 \phi^2(x)
\label{eq:Ham_scalar_Ham}
\end{align}
Observe that this is (semi-) positive definite.

The Hamiltonian formalism can be used as a stepping stone to impose
canonical quantization: we promote the fields
$\phi(t,\mathbf{x})$ and $\pi(t,\mathbf{x})$ to operators $\hat{\phi}_H(t,\mathbf{x})$
and $\hat{\pi}_H(t,\mathbf{x})$, respectively.
These are quantum fields (in the Heisenberg picture) 
which obey the canonical 
equal-time commutation relations:
\begin{equation}
\begin{aligned}
\left [ \hat{\phi}_H(t,\mathbf{x}), \hat{\pi}_H(t,\mathbf{y}) \right ] &= i \delta^{(3)} (\mathbf{x} - \mathbf{y})
 \\
\left [ \hat{\phi}_H(t,\mathbf{x}), \hat{\phi}_H(t,\mathbf{y}) \right ] &=
\left [ \hat{\pi}_H(t,\mathbf{x}), \hat{\pi}_H(t,\mathbf{y}) \right ] = 0
\label{eq:qft_CCR}
\end{aligned}
\end{equation}
Together with the Heisenberg equations of motion:
\begin{equation}
\begin{aligned}
i \frac{\partial}{\partial t}\hat{\phi}_H(t,\mathbf{x}) &= [\hat{\phi}_H(t,\mathbf{x}),\hat{H}]  \\
i \frac{\partial}{\partial t}\hat{\pi}_H(t,\mathbf{x}) &= [\hat{\pi}_H(t,\mathbf{x}),\hat{H}]
\label{eq:QFT_EOM}
\end{aligned}
\end{equation}
the commutation relations lead to the following field equation:
\begin{equation}
\left [ \frac{\partial^2}{\partial t^2} 
- \nabla^2 + m^2
\right ] \hat{\phi}_H(t,\mathbf{x}) = 0
\label{eq:qft_KG_oper}
\end{equation}
which is an operator version of the Klein--Gordon equation from Eq.~(\ref{eq:KG_ch2}). 
Crucially, this was a \textit{result}, not a mere promotion of the classical
field equation.
Its solution is, similarly, 
an operator version of the field expansion in Eq.~(\ref{eq:scalar_field_exp}), namely the Hermitian quantum field:
\begin{equation}
\hat{\phi}_H(t,\mathbf{x}) =  \int \frac{d^3k}{(2\pi)^3 2\omega} \left [ \hat{a}(\mathbf{k}) e^{-ikx}
+ \hat{a}^{\dagger}(\mathbf{k}) e^{ikx} \right ]
\label{eq:exp_Phi}
\end{equation}
where the time-independent coefficients $\hat{a}(\mathbf{k})$ and $\hat{a}^{\dagger}(\mathbf{k})$ are now operators. If we now plug this field expansion
back into the Hamiltonian, we will be pleased to find out that the latter is diagonal:
\begin{equation}
\hat{H} = \int \frac{d^3k}{(2\pi)^3 2\omega}
\omega  \hat{a}^{\dagger}(\mathbf{k}) \hat{a}(\mathbf{k})
\end{equation}
where we implicitly normal-ordered; the N-operation here (normal-ordering) places all creation operators
to the left of all annihilation operators.
Another important quantity to consider is the time-ordered two-point
function:
\begin{equation}
\Delta_F(x_A-x_B)  \equiv \langle 0 | T  [ \hat{\phi}_H(x_A) \hat{\phi}_H(x_B)  ] | 0 \rangle =
\int \frac{d^4k}{(2\pi)^4} \frac{i}{k^2 - m^2 + i \epsilon} e^{-ik(x_A-x_B)}
\label{eq:TO3}
\end{equation}
also known as a Feynman propagator. The T-operation means ``later to the left.''

The above approach (promote classical to quantum fields, impose equal-time commutation relations, use the Heisenberg equations of motion, and solve the field equation via a Fourier-mode decomposition) works fine for the non-interacting theory of Eq.~(\ref{eq:Ham_scalar_Ham}), but gets us in trouble as soon as we turn interactions on. In the latter case, a better approach is to work in the interaction picture,
splitting the Hamiltonian into an ``easy'' and a ``hard'' part:
\begin{equation}
\begin{aligned}
\hat{H}_0^I &= \int d^3 x \left ( \frac{1}{2} \hat{\pi}_I^2(t,\mathbf{x}) + \frac{1}{2} (\nabla \hat{\phi}_I(t,\mathbf{x}))^2   
+ \frac{1}{2} m^2 \hat{\phi}_I^2(t,\mathbf{x})  \right ) \\
\hat{H}_1^I(t) &= \frac{\lambda}{4} \int d^3 x \hat{\phi}_I^4(t,\mathbf{x})
\end{aligned}
\end{equation}
where the field expansion in Eq.~(\ref{eq:exp_Phi}) now applies
to the interaction-picture quantum field, $\hat{\phi}_I(t,\mathbf{x})$,
and we are studying the case of a quartic interaction.
The idea, then, is to take the Dyson series for the scattering
operator:
\begin{align}
\hat{S}
= \sum_{n=0}^{\infty} \left ( -i \right )^n \frac{1}{n!} 
\int d^4 x_1 \int d^4 x_2 
\cdots \int d^4 x_n 
T\left [ \hat{{\cal H}}_1^I(x_1) \hat{{\cal H}}_1^I(x_2) \cdots \hat{{\cal H}}_1^I(x_n) \right ] 
\label{eq:Dyson1}
\end{align}
and sandwich it between specific states, 
$S_{ba} \equiv \langle u_b | \hat{S} | u_a \rangle$, 
to produce the S-matrix (amplitude), 
which is related to experimental observables. 

Our task is
to evaluate the vacuum expectation value of time-ordered products
of increasingly more and more operators. This is vastly
simplified if we introduce the concept of a Wick contraction:
\begin{equation}
T [ \hat{\phi}_I(x) \hat{\phi}_I(y) ] =  N [ \hat{\phi}_I(x) \hat{\phi}_I(y) ]
+ \contraction{}{\hat{\phi}_I(x)}{}{\hat{\phi}_I(y)}
\hat{\phi}_I(x) \hat{\phi}_I(y)
\end{equation}
The many-operator generalization of the above is known as Wick's theorem:
\begin{align}
T[\hat{A} \hat{B} \hat{C} \hat{D}\hat{E} \hat{F} \cdots \hat{Z}] &= 
N[\hat{A} \hat{B} \hat{C} \hat{D}\hat{E} \hat{F} \cdots \hat{Z}] \nonumber \\
&\quad + N [ \contraction{}{\hat{A}}{}{\hat{B}}
\hat{A} \hat{B} \hat{C} \hat{D}\hat{E} \hat{F} \cdots \hat{Z}] 
+ N [ \contraction{\hat{A}}{\hat{B}}{\hat{C}}{\hat{D}}
\hat{A} \hat{B} \hat{C} \hat{D}\hat{E} \hat{F} \cdots \hat{Z}] + 
N [ \contraction{}{\hat{A}}{\hat{B}\hat{C}}{\hat{D}}
\hat{A} \hat{B} \hat{C} \hat{D}\hat{E} \hat{F} \cdots \hat{Z}] + \cdots \nonumber \\
&\quad + N [ \contraction{}{\hat{A}}{}{\hat{B}} 
\contraction{\hat{A}\hat{B}}{\hat{C}}{}{\hat{D}}
\hat{A} \hat{B} \hat{C} \hat{D}\hat{E} \hat{F} \cdots \hat{Z}]
+ N [ \contraction{}{\hat{A}}{\hat{B}}{\hat{C}} 
\contraction[2ex]{\hat{A}}{\hat{B}}{\hat{C}}{\hat{D}}
\hat{A} \hat{B} \hat{C} \hat{D}\hat{E} \hat{F} \cdots \hat{Z}]
+ N [ \contraction{\hat{A}}{\hat{B}}{}{\hat{C}} 
\contraction[2ex]{}{\hat{A}}{\hat{B}\hat{C}}{\hat{D}}
\hat{A} \hat{B} \hat{C} \hat{D}\hat{E} \hat{F} \cdots \hat{Z}] + \cdots \nonumber \\
&\quad + N [ \contraction{}{\hat{A}}{}{\hat{B}} 
\contraction{\hat{A}\hat{B}}{\hat{C}}{}{\hat{D}}
\contraction{\hat{A}\hat{B}\hat{C}\hat{D}}{\hat{E}}{}{\hat{F}}
\hat{A} \hat{B} \hat{C} \hat{D}\hat{E} \hat{F} \cdots \hat{Z}]
+ N [ \contraction{}{\hat{A}}{\hat{B}}{\hat{C}} 
\contraction[2ex]{\hat{A}}{\hat{B}}{\hat{C}}{\hat{D}}
\contraction{\hat{A} \hat{B} \hat{C} \hat{D}}{\hat{E}}{}{\hat{F}}
\hat{A} \hat{B} \hat{C} \hat{D}\hat{E} \hat{F} \cdots \hat{Z}]
+ N [ \contraction{\hat{A}}{\hat{B}}{}{\hat{C}} 
\contraction[2ex]{}{\hat{A}}{\hat{B}\hat{C}}{\hat{D}}
\contraction{\hat{A} \hat{B} \hat{C} \hat{D}}{\hat{E}}{}{\hat{F}}
\hat{A} \hat{B} \hat{C} \hat{D}\hat{E} \hat{F} \cdots \hat{Z}] + \cdots \nonumber \\
&\quad + N[\text{higher contractions}]
\label{eq:Wick}
\end{align}
This expresses the time-ordered product of a set of operators as the sum of contracted normal-ordered products of the same operators.

\begin{figure}[t]
\centering
   \begin{subfigure}[b]{0.24\textwidth} \centering
     \includegraphics[scale=0.08]{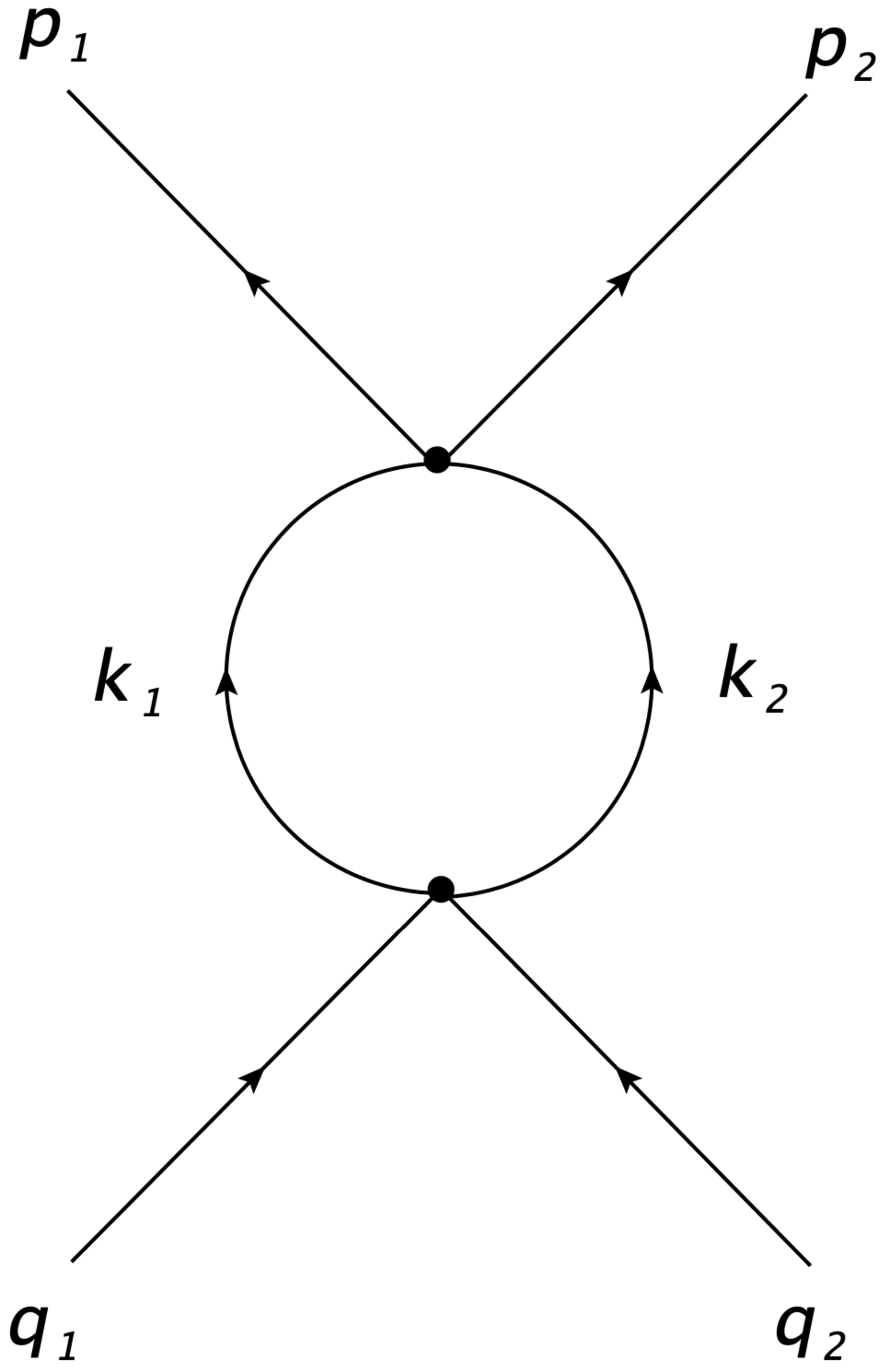}
     \caption{}
     \label{fig:Mom22_A2}
   \end{subfigure}
   \begin{subfigure}[b]{0.37\textwidth} \centering
     \includegraphics[scale=0.09]{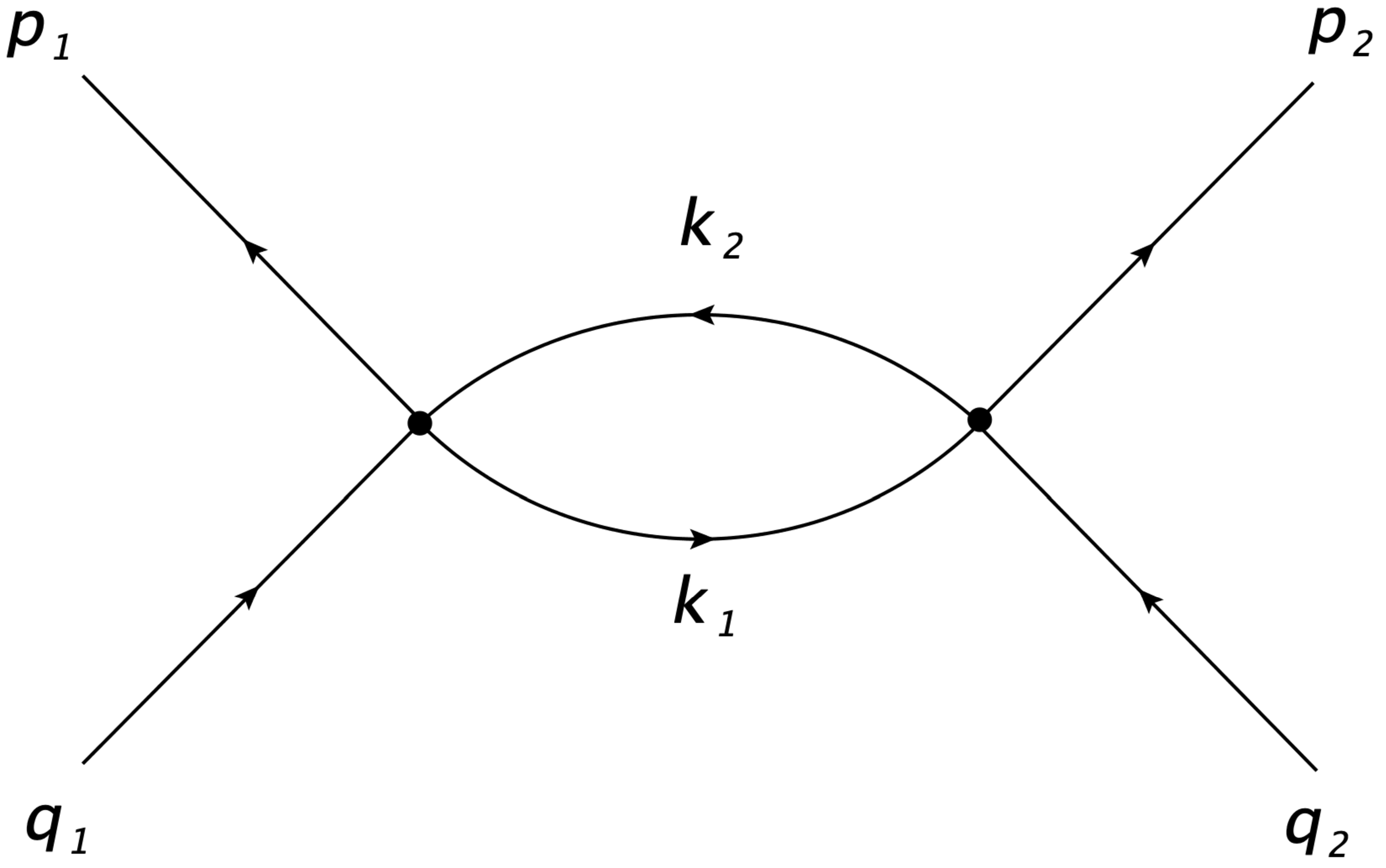
     }
     \caption{}
     \label{fig:Mom22_B2}
   \end{subfigure}
   \begin{subfigure}[b]{0.37\textwidth} \centering
     \includegraphics[scale=0.09]{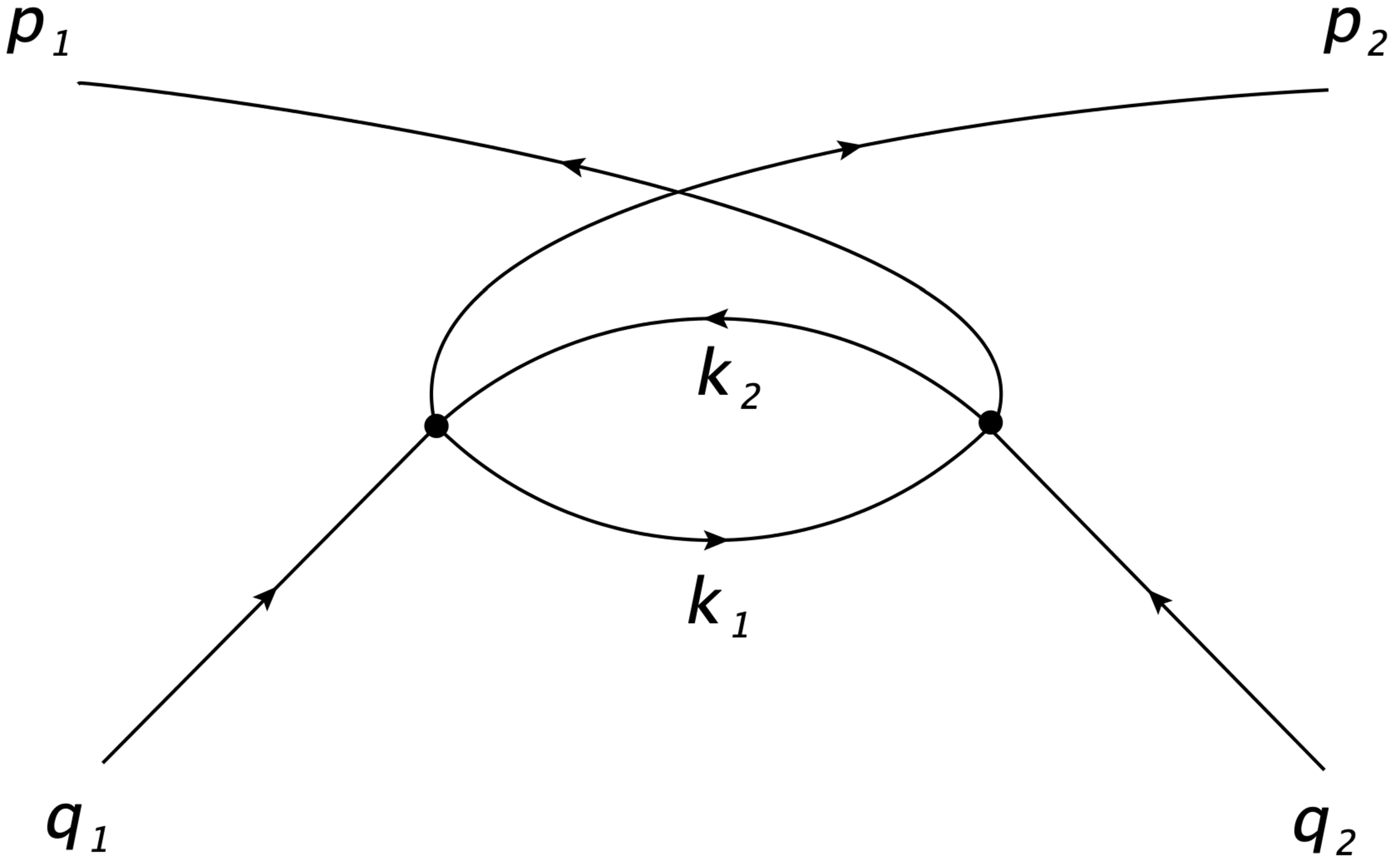}
     \caption{}
     \label{fig:Mom22_C2}
   \end{subfigure}

\caption{Connected, momentum-space, second-order Feynman diagrams representing the S-matrix in a 
$2 \rightarrow 2$ process (for a quartic interaction) in the $s$, $t$, and $u$ channels.}
\label{fig:Mom22}
\end{figure}

Consider a specific physical setting, where you have two particles
in the initial state and two in the final state. The S-matrix then becomes:
\begin{align}
&S_{\mathbf{p}_1\mathbf{p}_2, \mathbf{q}_1 \mathbf{q}_2} = \langle 0 | \hat{a}(\mathbf{p}_1) \hat{a}(\mathbf{p}_2) \hat{S}\hat{a}^{\dagger}(\mathbf{q}_1) \hat{a}^{\dagger}(\mathbf{q}_2) | 0 \rangle 
\nonumber \\
&=  \langle 0 | \hat{a}(\mathbf{p}_1)\hat{a}(\mathbf{p}_2) \hat{a}^{\dagger}(\mathbf{q}_1)\hat{a}^{\dagger}(\mathbf{q}_2) | 0 \rangle
+ (-i) \frac{\lambda}{4} 
\int d^4 x \langle 0 | T\left [ \hat{a}(\mathbf{p}_1)\hat{a}(\mathbf{p}_2) \hat{\phi}^4_I(x) \hat{a}^{\dagger}(\mathbf{q}_1)\hat{a}^{\dagger}(\mathbf{q}_2) \right ]  | 0 \rangle
\nonumber \\
&\quad + (-i)^2 \frac{1}{2!} \left( \frac{\lambda}{4}\right)^2 \int d^4 x d^4 y 
\langle 0 | T\left [ \hat{a}(\mathbf{p}_1) \hat{a}(\mathbf{p}_2) \hat{\phi}^4_I(x) \hat{\phi}^4_I(y) \hat{a}^{\dagger}(\mathbf{q}_1)\hat{a}^{\dagger}(\mathbf{q}_2) \right ] | 0 \rangle
+ \cdots
\label{eq:feyn_mom_4b}
\end{align}
which can be evaluated using Wick's theorem.
At second order, the contributions can be summarized graphically via
the (momentum-space) Feynman diagrams of Fig.~\ref{fig:Mom22}.
Symbolically, all three contributions take the form of the following
loop integral:
\begin{align}
J(l^2) &=  
\int \frac{d^4k}{(2\pi)^4}  \frac{1}{k^2 - m^2 + i \epsilon}  \frac{1}{(k - l)^2 - m^2 + i \epsilon}
\label{eq:feyn_mom_10}
\end{align}
where our notation reflects the fact that this quantity is Lorentz invariant.
We can combine the two denominators using the following
identity:
\begin{align}
\frac{1}{ab} = \int_0^1 \frac{dx}{[(1-x)a + xb]^2}
\label{eq:ccreno_2}
\end{align}
where $x$ is known as Feynman parameter. This leads to:
\begin{align}
J(l^2) &=  
\int_0^1 dx
\int \frac{d^4k}{(2\pi)^4} \frac{1}{\left [ k^2  - \Delta + i \epsilon \right ]^2}
\label{eq:feyn_mom_11}
\end{align}
where $\Delta \equiv m^2 - x(1-x) l^2$.

\section{Mistakes and corrections}

Just like in Refs.\ \onlinecite{GezerlisWilliams1} and \onlinecite{GezerlisWilliams2}, our goal is to discuss how certain
reasonably subtle topics 
should be correctly understood, not to criticize authors who toiled
hard to produce respected textbooks on a given subject. Thus,
we will now cite a superset of references, containing standard textbooks
that all discuss related 
topics.\cite{Aitchison,Alvarez,Banks,Baulieu,Bogoliubov,Brown,Coleman,Das,Donoghue,Folland,Fradkin,
Gelis,Greiner,Gross,Hatfield,Itzykson,Kleinert,Lancaster,Maggiore,Mandl,Nastase,Padmanabhan,Peskin,
Radovanovic,Ramond, 
Ryder,Schwartz,Srednicki,Stone,Talagrand,Williams,Zee,Zinn-Justin}  
(We have shared detailed bibliographic
information with the Editor and Referees of the present manuscript on specific instances from the literature  where the incorrect claims appear.) 
For each theme that is to follow, we first provide some context,
then give a specific quote from a QFT textbook (the \textit{mistake}), 
proceed to discuss how and why 
the physics going into the quote is wrong, and in the end provide a few-sentence improved version 
(the \textit{correction}).
To keep the coverage from ballooning, some of these misconceptions are touched upon 
but not explored in detail (after all, this work addresses not one, but \textit{six} mistakes); with that in mind, we cite works by acknowledged experts, where the reader can discover more on each subject.
In some of the excerpts we have tweaked the notation, in order to keep
things consistent with section~\ref{sec:notation}; the original 
quotes have not been modified in other ways (unless explicitly marked).

\subsection{Relativistic quantum mechanics}

Textbooks on quantum mechanics (QM) typically start with a brief quasi-historical overview of the old quantum theory. Similarly, textbooks on quantum field theory often have an introductory chapter on relativistic quantum mechanics, pointing out that it is difficult to set up a consistent theory that way, thereby motivating the alternative approach of quantum field theory. While that is reasonable enough, one should not go overboard: it doesn't help readers to
claim that an old approach is more diseased than it actually is. 
The methodological principle of charity is not merely a matter of 
principle: some of the alleged problems with relativistic QM re-appear in 
the context of QFT, but they are typically passed over in silence. A careful reader/student is then left with the discomfort of not fully understanding 
why one approach is bad and the other good.

{\bf Mistake \#1} ``\textit{Thus, negative energies 
$E = -\sqrt{\mathbf{p}^2 + m^2}$		
are on the same footing as the physical ones 
$E = +\sqrt{\mathbf{p}^2 + m^2}$. This is a severe difficulty 
because the spectrum is no longer bounded 
from below. It seems that an arbitrarily large
amount of energy may be extracted from the 
system […] This is clearly a failure of the 
concept of stable stationary states.}''

The claim here is that the negative energy solutions of the Klein--Gordon
equation, cf. Eq.~(\ref{eq:KG_ch2}), in its guise as a single-particle relativistic generalization of
the Schr\"odinger equation:
\begin{equation}
\left [ \frac{\partial^2}{\partial t^2} - \nabla^2 + m^2
\right ] \psi(t,\mathbf{x}) = 0
\end{equation}
are a conceptual blight, which forces us to stop
studying relativistic QM. To be explicit, this refers to the plane-wave
solutions:
\begin{equation}
\psi(t,\mathbf{x}) = N e^{i(\mathbf{p} \cdot \mathbf{x} - Et)} 
\label{eq:kg_plane}
\end{equation}
for which $E = \pm \sqrt{\mathbf{p}^2 + m^2}$. 

A connection is typically also made with the continuity equation:
\begin{equation}
\frac{\partial \rho}{\partial t} + \nabla \cdot \mathbf{j} = 0
\label{eq:contin_QM}
\end{equation}
where, for our problem:
\begin{equation}
\begin{aligned}
\rho &= \frac{i}{2m} \left ( \psi^*\frac{\partial \psi}{\partial t} - \psi\frac{\partial \psi^*}{\partial t} \right ),\quad
\mathbf{j} &= \frac{1}{2 m i} \left ( \psi^* \nabla \psi - \psi \nabla \psi^* \right )
\label{eq:rho_j_KG}
\end{aligned}
\end{equation}
If we plug in the plane waves of Eq.~(\ref{eq:kg_plane}), this gives:
\begin{equation}
\begin{aligned}
\rho = |N|^2 \frac{E}{m},\quad
\mathbf{j} = |N|^2 \frac{\mathbf{p}}{m} 
\label{eq:rho_in_KG}
\end{aligned}
\end{equation}
The alleged issue is that the 
probability density contains $E$ but, since
$E = \pm \sqrt{\mathbf{p}^2 + m^2}$ tells us that $E$ can be either positive or negative, $\rho$ 
\textit{isn't positive definite}.
Crucially, the conclusions on both $E$ and $\rho$ are drawn still
at the level of a single, non-interacting particle.

In classical mechanics, the existence of unphysical solutions is typically
sidestepped by noting that they can simply be omitted, precisely because
they are unphysical. 
Unlike what so many modern textbooks claim, exactly the same thing can be
done in a relativistic quantum mechanical theory.
Negative energy solutions are not an issue, if all we're dealing with 
is a free particle: 
a particle in a positive-energy state $E = + \sqrt{\mathbf{p}^2 + m^2}$
that doesn't experience any interactions will simply remain in that state.
Similarly, since $E>0$ we see from Eq.~(\ref{eq:rho_in_KG}) that 
$\rho$ will also remain positive at all times. As pointed out 
in the early classic QFT textbook by S. Schweber 
(Ref.\ \onlinecite{Schweber}, p. 56):
``a consistent theory can be developed for a free particle if we adopt the manifold of positive energy solutions as the set of states which are physically realizable by a free particle.''
A similar argument can be put forward in terms of wave packets, 
modifying Eq.~(\ref{eq:scalar_field_exp})---see also section~\ref{sec:local} below.

You may be experiencing minor discomfort at this stage: can one just
arbitrarily drop one half of our solutions? Isn't that conceptually unsatisfying?
In other words, perhaps one \textit{does} need to give up on relativistic QM in
favor of a more advanced theory (namely QFT) that doesn't suffer from this 
arbitrariness? The part of the story that the texbooks typically leave out
is that we are faced with a similar arbitrariness in field theory:
we mentioned above that when applying the current coming from
Noether’s theorem, Eq.~(\ref{eq:Noether_first_curr})
to the case of time translation, the conserved charge is precisely the Hamiltonian $H$. But why did we pick \textit{that} sign in the current of Eq.~(\ref{eq:Noether_first_curr}) in the first place? After all, 
if $\partial_{\mu} j^{\mu} = 0$ then $\partial_{\mu} (-j^{\mu}) = 0$ also holds.
The answer is simply that we wrote the Noether current in that specific 
form in order to ensure that we would end up with a positive definite Hamiltonian.
In other words, we \textit{chose} to work with a positive energy even in a field theoretic context.

{\bf Correction \#1} \textit{There is nothing wrong with using the Klein-Gordon equation to describe a single relativistic particle, provided there are no interactions/perturbations involved. A particle that starts in
a positive energy state will remain in that state. A similarly arbitrary choice in favor of positive energy is also made in field theory.
The relativistic QM story is, indeed, complicated if you turn on the interactions but, then again, interacting QFT isn't child's play, either.}

\subsection{Noether's theorem}

In section~\ref{sec:notation} we stated Noether's theorem in words
(every continuous global symmetry transformation leads to a conserved current)
and then showed the Noether current, Eq.~(\ref{eq:Noether_first_curr}), without
proof. This was given in the general case where both internal and spacetime symmetry transformations are being considered at the same time. Many standard
textbooks study these two scenarios separately (or even worse, examine
special cases without a general derivation), thereby obscuring the physics
behind this most important theorem. As we will now see, sometimes even
the sources that \textit{do} go into a general derivation of Noether's theorem 
use notation that ranges from impenetrable to flat-out wrong.

{\bf Mistake \#2} ``\textit{We now show how to derive [the field equation] from a variational principle applied to an action:
\begin{equation}
\mathcal{S} = \int \mathcal{L}(\phi, \partial_{\mu}\phi) d^4 x
\end{equation}
[\ldots] We now subject both the field variable and the coordinates to a variation
which vanishes on the boundary $\partial R$:
\begin{equation}
\begin{aligned}
x^{\mu} &\rightarrow {x'}^{\mu} = x^{\mu} + \delta x^{\mu},\nonumber \\
\phi(x) &\rightarrow \phi'(x) = \phi(x) + \delta\phi(x) 
\end{aligned}
\end{equation}
It is convenient to consider the case where $\mathcal{L}$ depends explicitly on
$x^{\mu}$:
\begin{equation}
\mathcal{L} = \mathcal{L}(\phi, \partial_{\mu}\phi, x^{\mu})
\end{equation}
this happens if $\phi$ interacts with an external source, and so does not describe
a closed system.}''

Let us briefly summarize what is at stake: like all textbooks studying
scalar QFT, this one makes the reasonable assumption that the main focus
of study will be theories of the form $\mathcal{L}(\phi, \partial_{\mu}\phi)$---e.g., a theory made up of a kinetic term and a quartic 
self-interaction. However, instead of making that assumption and sticking
to it when the time comes to derive Noether's theorem, this introductory
textbook feels the need to soften the requirement 
of translation invariance (which forbids an explicit $x^{\mu}$ dependence)
for a single section, only to return to the usual $\mathcal{L}(\phi, \partial_{\mu}\phi)$ later on. Similarly, other references go even farther when deriving
Noether's theorem, 
writing the Lagrangian density simply
as $\mathcal{L}(x)$, without any reference to fields.

The discomfort of these authors arises when they try to generalize the 
intrinsic and total variations of Eq.~(\ref{eq:intrinsic_first}) and Eq.~(\ref{eq:change_total}) to the Lagrangian density, namely:
\begin{equation}
{\cal L}(\phi'(x'), \partial \phi'(x')/\partial x'^{\mu}) = {\cal L}(\phi(x), \partial \phi(x)/\partial x^{\mu}) + \tilde{\delta} {\cal L}
\label{eq:change_total_Lag}
\end{equation}
and
\begin{equation}
{\cal L}(\phi'(x), \partial \phi'(x)/\partial x^{\mu}) = {\cal L}(\phi(x), \partial \phi(x)/\partial x^{\mu}) + \delta {\cal L}
\label{eq:change_intrinsic_Lag}
\end{equation}
The crucial part in the argument emerges when one tries to relate
the two variations via the formula:
\begin{equation}
\tilde{\delta} {\cal L} = \delta {\cal L} + \frac{\partial {\cal L}}{\partial x^{\mu}} \delta x^{\mu}
\label{eq:change_lag_conn}
\end{equation}
The issue has to do with the presence of the $\partial \mathcal{L}/\partial x^{\mu}$ term: if there is no explicit dependence on $x^{\mu}$, one would
naively think that the derivative vanishes, in which case the two
variations coincide. Hence, some authors choose to introduce an \textit{ad hoc}
$x$-dependence just to provide surface respectability to Eq.~(\ref{eq:change_lag_conn}).

The resolution comes from considering the Lagrangian density without
any extraneous assumptions:
${\cal L} = {\cal L}(\phi(x), \partial \phi(x)/\partial x^{\nu})$.
In Eq.~(\ref{eq:change_lag_conn}), we are faced with the chain rule
in partial differentiation, when there are four independent variables (the $x^{\mu}$'s):
\begin{equation}
\frac{\partial {\cal L}(\phi(x), \partial \phi(x)/\partial x^{\nu})}{\partial x^{\mu}} = \frac{\partial {\cal L}(\phi(x), \partial \phi(x)/\partial x^{\nu})}{\partial \phi} \frac{\partial \phi(x)}{\partial x^{\mu}} + \frac{\partial {\cal L}(\phi(x), \partial \phi(x)/\partial x^{\nu})}{\partial (\partial \phi(x)/\partial x^{\nu})} \frac{\partial^2 \phi(x)}{\partial x^{\nu} \partial x^{\mu}}
\end{equation}
as shown (though not emphasized) on p. 12 of the very
careful textbook by G. Sterman.\cite{Sterman} 
If you're thinking that one should have been using the total derivative, 
$d / d x^{\mu}$, here---or, say, in the Euler-Lagrange Eq.~(\ref{eq:scalar_Lag})---you will be disappointed:
that notation is relevant only when there is a \textit{single} independent variable, whereas we are dealing with four in $x^{\mu}$.
According to Salam’s criterion,\cite{Salam} one's goal is always ``to find a notation which is both concise and intelligible to at least two people of whom one may be the author.''

{\bf Correction \#2} \textit{When dealing with a theory $\mathcal{L}(\phi, \partial_{\mu}\phi)$, we don’t get to introduce an explicit $x$-dependence by hand just because we feel uncomfortable about the presence of the derivative      
$\partial \mathcal{L}/\partial x^{\mu}$ in the derivation of Noether's theorem.
The notation reflects the chain rule in partial differentiation. There is no need for an explicit $x$-dependence and that's a good thing, because Noether's theorem
applies to translation-invariant theories.}

\subsection{Lagrangians and canonical quantization}

Our crash course on QFT in section~\ref{sec:notation} followed
a pretty standard route: classical field theory in the Lagrangian
formalism, a transition to the Hamiltonian formalism (still at the classical level),
and then canonical quantization (using the Hamiltonian formalism and promoting
classical fields to operators). In our exposition we were clear that
the Euler--Lagrange equations are a classical (and Lagrangian-based) construct,
whereas the Heisenberg equations of motion were a quantum (and Hamiltonian-based) construct.
While the result in both cases was the same (the Klein--Gordon equation, for 
a non-interacting theory), the flavor of the two arguments was very different.
The narrative is complicated by authors who, without much fanfare, go on to 
discuss a linear combination of the two approaches: 

{\bf Mistake \#3} ``\textit{The Hamiltonian and Lagrangian density operators have the same relationship as their classical counterparts and are related by a Legendre transformation,
\begin{equation}
\hat{H} \equiv \int d^3 x \hat{\mathcal{H}} \equiv 
\int d^3 x \hat{\pi}(x) \dot{\hat{\phi}}(x) - \hat{L}
= \int d^3 x \left ( \hat{\pi}(x) \dot{\hat{\phi}}(x) - \hat{{\cal L}} \right )
\end{equation}
where the canonical momentum density operator $\hat{\pi}(x)$ is defined as
\begin{equation}
\hat{\pi}(x) \equiv \frac{\delta \hat{L}}{\delta \dot{\hat{\phi}}(x)}
= \frac{\partial \hat{{\cal L}}}{\partial \dot{\hat{\phi}}(x)}
\end{equation}
Since the operator equations of motion are the same as their classical equivalents, then the operators
must also obey the Euler-Lagrange equations [\ldots] at the operator level for the Heisenberg
picture operator $\hat{\phi}(x)$, [\ldots]
\begin{equation}
\partial_{\mu} \frac{\partial \hat{{\cal L}}}{\partial 
(\partial_{\mu} \hat{\phi})(x)} - \frac{\partial \hat{{\cal L}}}{\partial \hat{\phi}(x)} = 0
\end{equation}
}''

This quote (and many others like it) is mixing the two general
philosophies which we were careful to distinguish above: in canonically
quantized QFT, quantum fields appear only in the Hamiltonian formalism, 
while the Lagrangian formalism involves only classical fields. 
(Admittedly, Schwinger's quantum action formalism---see section 2.1 of P. Roman's 
unjustly forgotten textbook\cite{Roman}---indeed combines the two,
but does so consistently, unlike discussions which 
promote classical to quantum fields
in selected equations \textit{by fiat}. Most notably, such an approach
motivates the equal-time commutation relations of Eq.~(\ref{eq:qft_CCR}).)
Thus, the fact that one ends up with the same field equation is a
\textit{result}, whereas in the quote this is 
essentially a starting assumption. 

The main issue here is that it is far from clear why
one should abandon the clear path discussed above (and in all QFT textbooks) for 
a mixing approach if one doesn't introduce any added benefits.
At a more detailed level, 
there are three main reasons why it is ill-advised for one to 
promote classical to quantum fields in the Lagrangian formalism,  
thereby ending up with a Lagrangian density which is an operator,
$\hat{{\cal L}}$. First, there is the principle of the matter:
as S. Weinberg points out on p. 300 of his classic textbook,\cite{Weinberg}
in all our theories we require that the action $\mathcal{S}$ be real.
If the Lagrangian density is an operator, then its spacetime integral
(the action) is also an operator, but then it cannot be a real number.
Second, in the path-integral approach to quantum field theory 
(not further touched upon in this article) one is faced
with the action and quantum fields that are 
c-number functions (not operators). Thus, 
it is very confusing to beginners, who are still trying to figure out
what is an operator and what is not, to be told that 
quantum fields show up as operators in a Lagrangian context,
despite the fact that we only talk about Lagrangians before we start the canonical
quantization program (and quantum fields are not operators in the path integral
program).

Third, one must be careful in handling
classical vs quantum fields \textit{even when} studying the Hamiltonian
formalism in the canonical quantization setting. The prototypical case
in this connection is that of derivative couplings; let's take 
the theory of scalar electrodynamics:
\begin{equation}
{\cal L} = \partial_{\mu} \varphi^*\partial^{\mu} \varphi 
- m^2 \varphi^*\varphi
- \frac{1}{4} F_{\mu \nu} F^{\mu \nu}
- iq(\varphi^*\partial^{\mu} \varphi
- \varphi \partial^{\mu} \varphi^*) A_{\mu} 
+ q^2 A_{\mu}A^{\mu} \varphi^* \varphi 
\label{eq:proj00}
\end{equation}
involving a complex scalar field and a gauge field.
The corresponding canonically conjugate momentum densities are: 
\begin{equation}
\varpi = \dot{\varphi}^* - iq \varphi^*A^0,
\qquad \varpi^* =  \dot{\varphi} + iq \varphi^*A^0,
\label{eq:proj0}
\end{equation}
Crucially, you can't willy-nilly promote classical fields to operators here.
Instead, one must employ the equation of motion
for an operator $\hat{O}_I(t)$ in the interaction picture, namely:
\begin{equation}
i\frac{\partial}{\partial t} \hat{O}_I(t) = [\hat{O}_I(t),\hat{H}_0^I]
\label{eq:interb_4}
\end{equation}
This leads to:
\begin{equation}
\hat{\varpi}_I(t,\mathbf{x}) = \frac{\partial}{\partial t}\hat{\varphi}_I^{\dagger}(t,\mathbf{x}), \qquad
\hat{\varpi}_I^{\dagger}(t,\mathbf{x}) = \frac{\partial}{\partial t}\hat{\varphi}_I(t,\mathbf{x})
\label{eq:proj5}
\end{equation}
which clearly look different than what we had in Eq.~(\ref{eq:proj0}).

{\bf Correction \#3} \textit{When canonically quantizing field theory, there is no need to promote fields to operators in the Lagrangian formalism. This helps you avoid ending up with an action that is not
a real number. It is much more natural to always start with the Lagrangian density,
move to the Hamiltonian formulation, and \textit{then} impose canonical quantization.}

\subsection{Particle localization}
\label{sec:local}

The question of what a quantum field \textit{really is} is left 
far too vague in far too many introductory treatments. Students
spend a good chunk of their undergraduate education learning 
that quantum mechanics is different from classical mechanics, only
to be told upon taking graduate QFT that everything is a quantum field
and particles (and the associated mechanics) are just the associated excitations.
This sounds like a pretty important point, but given the complexity of the calculations involved (as well as the need to study scalar, fermionic, and gauge fields), the core ideas in a QFT course 
are typically given short shrift. It is therefore
commendable that some authors employ analogies with non-relativistic
physics in order to put forward an interpretation of the quantum field; unfortunately, as we will soon see, 
an interpretation resulting from such an analogy is misleading.

{\bf Mistake \#4} ``\textit{
\begin{equation}
\left [ \hat{\phi}_S(\mathbf{x}), \hat{\phi}_S(\mathbf{y}) \right ] =
\left [ \hat{\pi}_S(\mathbf{x}), \hat{\pi}_S(\mathbf{y}) \right ] = 0
\label{eq:local1}
\end{equation}
(For now we work in the Schr\"odinger picture where $\phi$ and $\pi$ do not
depend on time.) [\ldots]
Finally let us consider the interpretation of the state $\hat{\phi}_S(\mathbf{x}) | 0 \rangle$. From the expansion
\begin{equation}
\hat{\phi}_S(\mathbf{x}) =  \int \frac{d^3k}{(2\pi)^3 2\omega} \left [ \hat{a}(\mathbf{k}) e^{i \mathbf{k}\cdot \mathbf{x}}
+ \hat{a}^{\dagger}(\mathbf{k}) e^{-i \mathbf{k}\cdot \mathbf{x}} \right ]
\label{eq:local2}
\end{equation}
we see that
\begin{equation}
\hat{\phi}_S(\mathbf{x}) | 0 \rangle = 
 \int \frac{d^3k}{(2\pi)^3 2\omega} e^{-i \mathbf{k}\cdot \mathbf{x}} 
| \mathbf{k} \rangle
\label{eq:local3}
\end{equation}
is a linear superposition of single-particle states that have well-defined momentum. Except for the factor $1/2\omega$, this is the same as the familiar nonrelativistic expression for the eigenstate of position $| \mathbf{x} \rangle$); in fact the extra factor is nearly constant for small (nonrelativistic) $\mathbf{k}$. We will therefore put forward the same interpretation, and claim that the operator 
$\hat{\phi}_S(\mathbf{x})$, acting on the vacuum, \text{creates a particle at position} $\mathbf{x}$.}''

As just mentioned, the motivation behind trying to interpret 
$\hat{\phi}_S(\mathbf{x}) | 0 \rangle$ is excellent. The operator
$\hat{a}^{\dagger}(\mathbf{k})$ acting on the vacuum 
creates a state $| \mathbf{k} \rangle$, 
so it is worthwhile to investigate what the effect of $\hat{\phi}_S(\mathbf{x})$ 
is. We can 
see from Eq.~(\ref{eq:local3}) that the right-hand side integrates
over single-particle states $| \mathbf{k} \rangle$, so it is quite
reasonable to assume that the left-hand side is a single-particle state itself. The problems
arise when authors take an extra step, ignoring the $1/2\omega$ in the denominator,
thereby concluding that $\hat{\phi}_S(\mathbf{x}) | 0 \rangle$ is
a particle \textit{at position} $\mathbf{x}$.

Particle localization in relativistic theories is more complicated than
that.\cite{Schweber} To give a flavor of what's involved, let us introduce
the Newton--Wigner position operator and a new localized field operator:
\begin{equation}
\hat{\mathbf{x}}_{nw} \equiv -i \nabla_{\mathbf{k}} + \frac{i}{2} \frac{\mathbf{k}}{\mathbf{k}^2 + m^2}, \qquad
\hat{\phi}_L(\mathbf{x}) \equiv 
\int \frac{d^3k}{(2\pi)^3} \frac{1}{\sqrt{2 \omega}} e^{i\mathbf{k} \cdot \mathbf{x}} \hat{a}(\mathbf{k}) 
\label{eq:NewtonWigner}
\end{equation}
Crucially, $\hat{\phi}_L(\mathbf{x})$ contains only a single plane-wave contribution.
It is straightforward to see that these two operators work together:
\begin{align}
\hat{\mathbf{x}}_{nw} \langle 0 | \hat{\phi}_L(\mathbf{y}) | \mathbf{k} \rangle
= \mathbf{y}  \langle 0 | \hat{\phi}_L(\mathbf{y}) | \mathbf{k} \rangle
\end{align}
to take the form of a position-eigenvalue equation. Crucially,
this new operator obeys the commutation relation:
\begin{equation}
[\hat{\phi}_L(\mathbf{x}), \hat{\phi}_L^{\dagger}(\mathbf{y})] = \delta^{(3)}(\mathbf{x} - \mathbf{y})
\end{equation}
which is clearly that of creation and annihilation operators in coordinate
space. One can certainly not say as much about the $\left [ \hat{\phi}_S(\mathbf{x}), \hat{\phi}_S(\mathbf{y}) \right ] = 0$ of Eq.~(\ref{eq:local1}).

It is also worthwhile in this regard to examine the relationship between
$\hat{\phi}_L(\mathbf{x})$ and the usual quantum field $\hat{\phi}_S(\mathbf{x})$.
One can combine Eq.~(\ref{eq:local2}) and the corresponding relationship
giving $\hat{\pi}_S(\mathbf{x})$ in terms of $\hat{a}(\mathbf{k})$
and $\hat{a}^{\dagger}(\mathbf{k})$, thereby deriving an equation which
gives $\hat{a}(\mathbf{k})$ in terms of $\hat{\phi}_S(\mathbf{x})$
and $\hat{\pi}_S(\mathbf{x})$. Plugging that result into Eq.~(\ref{eq:NewtonWigner})
gives:
\begin{align}
\hat{\phi}_L(\mathbf{x})  = \frac{1}{\sqrt{2}}\int d^3 y \hat{\phi}_S(\mathbf{y}) \int \frac{d^3k}{(2\pi)^3}  e^{i \mathbf{k} \cdot (\mathbf{x}-\mathbf{y})} 
\left [ \mathbf{k}^2 + m^2 \right ]^{1/4} 
 + i\sqrt{\frac{1}{2}}  \int d^3 y \hat{\pi}_S(\mathbf{y}) \int \frac{d^3k}{(2\pi)^3}  e^{i \mathbf{k} \cdot (\mathbf{x}-\mathbf{y})} \left [ \mathbf{k}^2 + m^2 \right ]^{-1/4}
\label{eq:phi_ptcle3}
\end{align}
If we do the integrals over $\mathbf{k}$, we will find 
some modified Bessel functions. The essential point here is that 
$\hat{\phi}_L(\mathbf{x})$ is a spatial integral of 
$\hat{\phi}_S(\mathbf{y})f(|\mathbf{x}-\mathbf{y}|)$ and $\hat{\pi}_S(\mathbf{y})g(|\mathbf{x}-\mathbf{y}|)$, namely  
a non-local function of the quantum fields; at large $|\mathbf{x}-\mathbf{y}|$, the 
function $f(|\mathbf{x}-\mathbf{y}|)$ 
goes as $\exp(-m|\mathbf{x}-\mathbf{y}|)/|\mathbf{x}-\mathbf{y}|^{9/4}$, 
while $g(|\mathbf{x}-\mathbf{y}|)$
goes as $\exp(-m|\mathbf{x}-\mathbf{y}|)/|\mathbf{x}-\mathbf{y}|^{7/4}$.
As R. Haag (of Haag's theorem fame) says on p. 33 of his monograph:\cite{Haag}
``for a massive particle the ambiguity in defining the localization is small, namely of the order of the Compton wavelength.'' 
As we will now see, small is quite different from zero. If 
in Eq.~(\ref{eq:phi_ptcle3}) we
take $|\mathbf{k}| \ll m$,
we can drop the $\mathbf{k}^2$ inside the square brackets,
so both integrals give a $\delta^{(3)} (\mathbf{x}-\mathbf{y})$,
allowing us to carry out the integration over $\mathbf{y}$. Then,
the operator $\hat{\phi}_L^{\dagger}(\mathbf{x})$
would be a \textit{local} function of $\hat{\phi}_S(\mathbf{x})$ and $\hat{\pi}_S(\mathbf{x})$. Thus, non-relativistically 
$\hat{\phi}_L(\mathbf{x})$ \textit{does}
correspond to a particle at a fixed position $\mathbf{x}$, but things
are different in the general case, where we care about Lorentz (rather
than Galilean) invariance.

{\bf Correction \#4} \textit{The quantum field corresponds
to a fixed position $\mathbf{x}$ only in the non-relativistic problem. In the relativistic
case, one needs to introduce a Newton--Wigner position operator as well as
a new operator $\phi_L^{\dagger}(\mathbf{x})$ creating particles at a fixed position $\mathbf{x}$, which does not coincide with the usual quantum field $\phi_S(\mathbf{x})$.}

\subsection{Wick's theorem vs normal-ordering}

In a typical textbook treatment, the tool of normal ordering
is first introduced when discussing the energy of the vacuum, for a non-interacting theory: a simple re-arrangement leads to the elimination of the zero-point motion.
Of course, interactions are (rightly) at the heart of any textbook treatment of 
quantum field theory. The N-operation re-appears in that context, as part of 
Wick's theorem,
which re-expresses a T-product (needed for the Dyson expansion giving us the 
S-matrix) as a sum of N-products. What's often lost in this discussion is
whether we should be normal-ordering the entire Hamiltonian (and if not, why). 
As part of the next mistake, we have therefore combined quotes from earlier and
later parts of a texbook (corresponding to non-interacting and interacting QFT,
respectively):

{\bf Mistake \#5} ``\textit{Congratulations, you are now the proud owner of a working quantum field theory, provided you remember the normal ordering interpretation. [\ldots]
Wick’s theorem can be illustrated for the case of four operators [\ldots]
In particular
\begin{equation}
\langle 0 | T [\hat{\phi}_I(x_1)\hat{\phi}_I(x_2)\hat{\phi}_I(x_3)\hat{\phi}_I(x_4)] | 0 \rangle = 
 \Delta_F(x_1-x_2)~\Delta_F(x_3-x_4) 
 + 
 \Delta_F(x_1-x_3)~\Delta_F(x_2-x_4)  
 +
  \Delta_F(x_1-x_4)~\Delta_F(x_2-x_3)
\end{equation}
where we have used the Feynman propagator $\Delta_F(x_1-x_2) = \langle 0 | T  [ \hat{\phi}_I(x_1) \hat{\phi}_I(x_2)  ] | 0 \rangle$.
[\ldots]
Using Wick’s theorem on the string 
$\langle 0 | \hat{a}(\mathbf{k}) \hat{\phi}^4(x) \hat{a}^{\dagger}(\mathbf{k}) | 0 \rangle = \langle 0 | \hat{a}(\mathbf{k}) \hat{\phi}(x)\hat{\phi}(x)\hat{\phi}(x)\hat{\phi}(x) \hat{a}^{\dagger}(\mathbf{k}) | 0 \rangle$, will yield up two sorts of term.
}

There is much to unpack here. First, we are told that we need to normal-order
the Hamiltonian. Then, we see an application of Wick's theorem, for a 
case where the time-ordered product of quantum fields is evaluated at four different
positions. This is fine so far as it goes, but the next excerpt shows an
interaction of the type $\hat{\phi}^4(x)$, which according to the previous
excerpt would lead to Feynman propagators evaluated at zero argument $\Delta_F(0)$.
These diverge, but that's not the main issue: the question is that 
the toy application of Wick's theorem is given for \textit{different} positions, whereas
the actual interaction involves multiple quantum fields evaluated at the \textit{same} position. These give rise to bubbles (also known as tadpoles):
unlike the Feynman diagrams of Fig.~\ref{fig:Mom22}, for bubbles 
a single initial particle and a single final particle are associated
with a loop. 

We noted above that normal-ordering a non-interacting Hamiltonian
eliminates the divergences associated with the 
zero-point motion; normal-ordering the interaction only
eliminates \textit{some} divergences. But there is a deeper reason
we should normal-order the interaction, one that is rarely
discussed in textbook treatments (p. 158 of Ref.\ \onlinecite{Duncan} being one of
the very few exceptions): the interaction term in a Hamiltonian
\textit{needs} to be normal-ordered if the cluster decomposition principle
(a pillar of the QFT edifice) is to be respected. In other words,
the string appearing in the last excerpt should, strictly speaking, 
not be $\langle 0 | \hat{a}(\mathbf{k}) \hat{\phi}^4(x) \hat{a}^{\dagger}(\mathbf{k}) | 0 \rangle$, but $\langle 0 | \hat{a}(\mathbf{k}) N[\hat{\phi}^4(x)] \hat{a}^{\dagger}(\mathbf{k}) | 0 \rangle$. 

As a perceptive reader may be deducing, we are going to need
a new version of Wick's theorem for
what are known as \textit{mixed} T-products, i.e., T-products which contain some (or all)
terms in normal-products, e.g.  $T[\hat{A} ~N[ \hat{B} \hat{C} \hat{D} ] ~ \hat{E} \cdots \hat{Z}]$,
where $\hat{B}$, $\hat{C}$, and $\hat{D}$ have the same time label.  
As it so happens, G. Wick was well aware of this issue already when 
proposing his (now) eponymous theorem:\cite{Wick} 
he put forward a ``Theorem 2,'' precisely to handle this situation. 
Wick's trick was to interpret 
$T[\hat{A} ~N[ \hat{B} \hat{C} \hat{D} ] ~ \hat{E} \cdots \hat{Z}]$
as the limit of $T[\hat{A} \hat{B} \hat{C} \hat{D} \hat{E} \cdots \hat{Z}]$ when  the creation
operators among $\hat{B}$, $\hat{C}$, and $\hat{D}$ have a time label that is infinitesimally later than  
the time label of the annihilation operators among $\hat{B}$, $\hat{C}$, and $\hat{D}$.
Basically, he rewrote equal-time normal-ordered products in terms of unequal-time
non-normal-ordered products. This automatically implies that contractions between equal-time normal-ordered operators vanish,

This version of Wick's theorem (sometimes called \textit{Wick's corollary})
is most easily grasped via an example:
\begin{align}
T \left [ \hat{\phi}_I(x) N \left [ \hat{\phi}_I(y) \hat{\phi}_I(y) \right ] \right ]
&= N \left [ \hat{\phi}_I(x) \hat{\phi}_I(y) \hat{\phi}_I(y) \right ]
+ N \left [ 
\contraction{}{\hat{\phi}_I(x)}{}{\hat{\phi}_I(y)}
\hat{\phi}_I(x) \hat{\phi}_I(y) \hat{\phi}_I(y) \right ]
 + N \left [ 
\contraction{}{\hat{\phi}_I(x)}{\hat{\phi}_I(y)}{\hat{\phi}_I(y)}
\hat{\phi}_I(x) \hat{\phi}_I(y) \hat{\phi}_I(y) \right ]
\label{eq:cor1}
\end{align}
What this is showing is the application of 
Wick's theorem 
to a mixed T-product of equal-time normal-ordered operators, with the crucial
proviso that contractions between factors that were already in normal product form (i.e., had the same time label) are being omitted
on the right-hand side: there is no contraction here
connecting $y$ with $y$. This \textit{ipso facto}
eliminates all bubbles from consideration. Intriguingly, the modified
version of Wick's theorem was typically addressed in early standard
textbooks, e.g., see p. 184 of the classic work by Bjorken \& Drell.\cite{Bjorken}

{\bf Correction \#5} \textit{One must normal-order the interaction, 
with a view to respecting the cluster decomposition
principle. In order to handle
normal-ordered interactions involving multiple quantum fields at the same position, one needs a modified version of Wick's theorem, wherein contractions that 
involve terms that were already normal-ordered are omitted. This eliminates
all bubbles.}

\subsection{Wick rotation}

Given how much material needs to fit into a one-semester course on QFT,
students are sometimes disappointed to find out that all the machinery
on the S-matrix, Feynman diagrams, etc. doesn't actually lead to any 
practical conclusions until the following semester. 
Regardless of when they are introduced, calculations like that leading
up to Eq.~(\ref{eq:feyn_mom_11}) are very important, because
they go beyond just writing down an integral and onto the question
of what the result actually is. A crucial step in that process
is the so-called ``Wick rotation''; incidentally, this is a misnomer,
since F. Dyson introduced this idea years before G. Wick.\cite{Dyson,Wick2} 
The specific argument involved will be discussed below, but qualitatively,  
the main takeaway is that one can trade Minkowski 4-vectors for
Euclidean 4-vectors.

{\bf Mistake \#6} ``\textit{By applying the Feynman parametrization, the integral becomes
\begin{align}
I =  
\int_0^1 dx
\int \frac{d^4k}{(2\pi)^4} \frac{1}{\left [ (l+px)^2  - \Delta \right ]^2}
\label{eq:wickrot1}
\end{align}
where $\Delta \equiv m^2 - x(1-x) l^2$. By making the change of variable
$k = l + px$ and going to Euclidean space ($k^0 = i k^0_E$, $\mathbf{k} = \mathbf{k}_E$) we get
\begin{equation}
I =  i \int_0^1 dx
\int \frac{d^4k_E}{(2\pi)^4} \frac{1}{\left ( k_E^2 + \Delta \right )^2}
\label{eq:wickrot2}
\end{equation}}''

There are two sub-fallacies here, both quite widespread. First, 
we are told that the transition to Euclidean 
space is a simple \textit{change of variables}:
$k^0 = i k^0_E$, $\mathbf{k} = \mathbf{k}_E$. To see why this is wrong, 
let us spell out the integration measure and the square of the Minkowski 4-vector 
in the denominator of the integrand in Eq.~(\ref{eq:feyn_mom_11}):
\begin{align}
J(l^2) &=  
\int_0^1 dx
\int \frac{d^3 k}{(2\pi)^4} \int_{-\infty}^{+\infty} dk^0 \frac{1}{\left [ (k^0)^2 - \mathbf{k}^2 - \Delta + i \epsilon \right ]^2}
\label{eq:wickrot3}
\end{align}
If we carried out the change of variables $k^0 = i k^0_E$ here, the 
denominator would look prettier, but the integration limits would not:
we would be stuck with imaginary integration limits, $\int_{-i\infty}^{+i\infty} dk_E^0$. In other words, we'd have a nicely symmetric sum of two
positive terms in the denominator, $(k_E^0)^2 + \mathbf{k}^2$, but the integration
over $\mathbf{k}$ would be in real space whereas that over $k_E^0$ would 
be over imaginary values. Nothing gained.

\begin{figure}[t]
\centering
   \begin{subfigure}[b]{0.44\textwidth} \centering
     \includegraphics[scale=0.41]{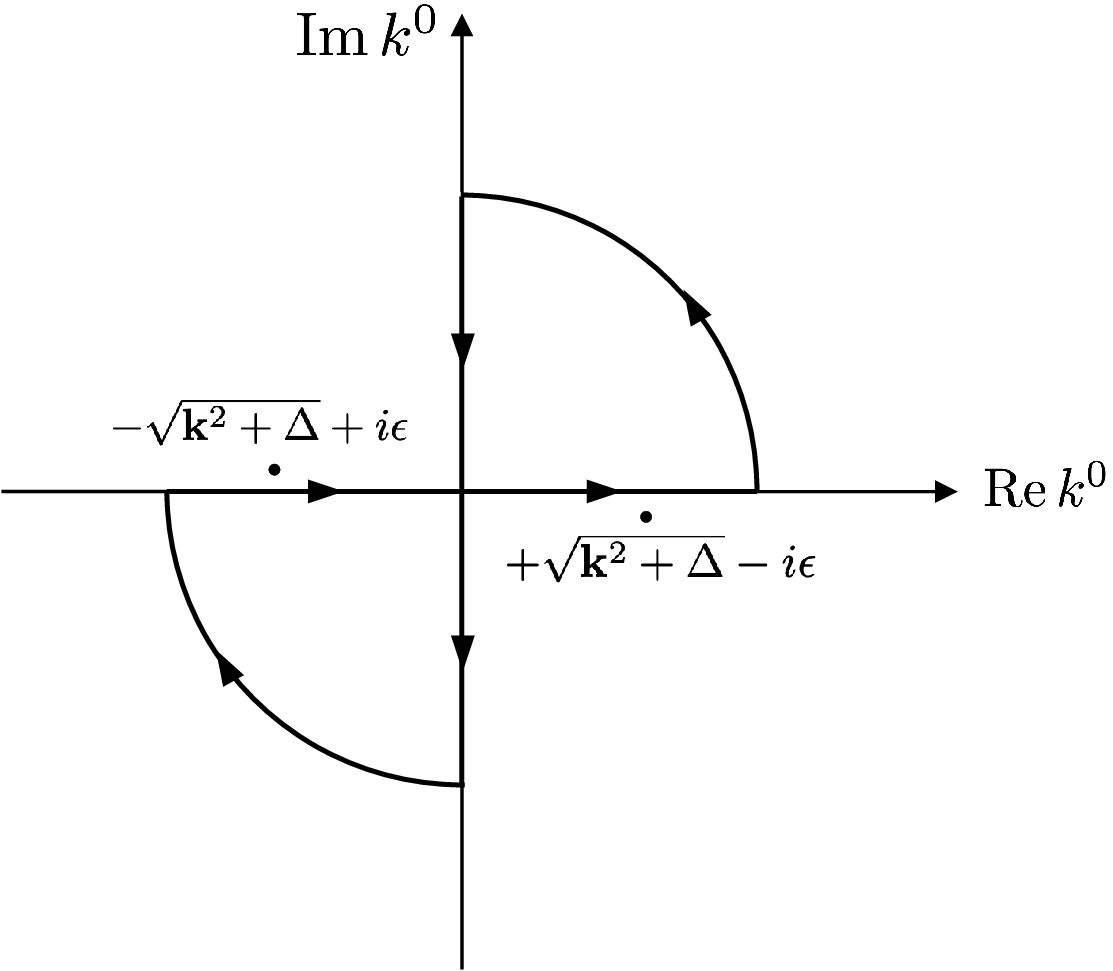}
     \caption{}
   \end{subfigure}
   \begin{subfigure}[b]{0.44\textwidth} \centering
     \includegraphics[scale=0.41]{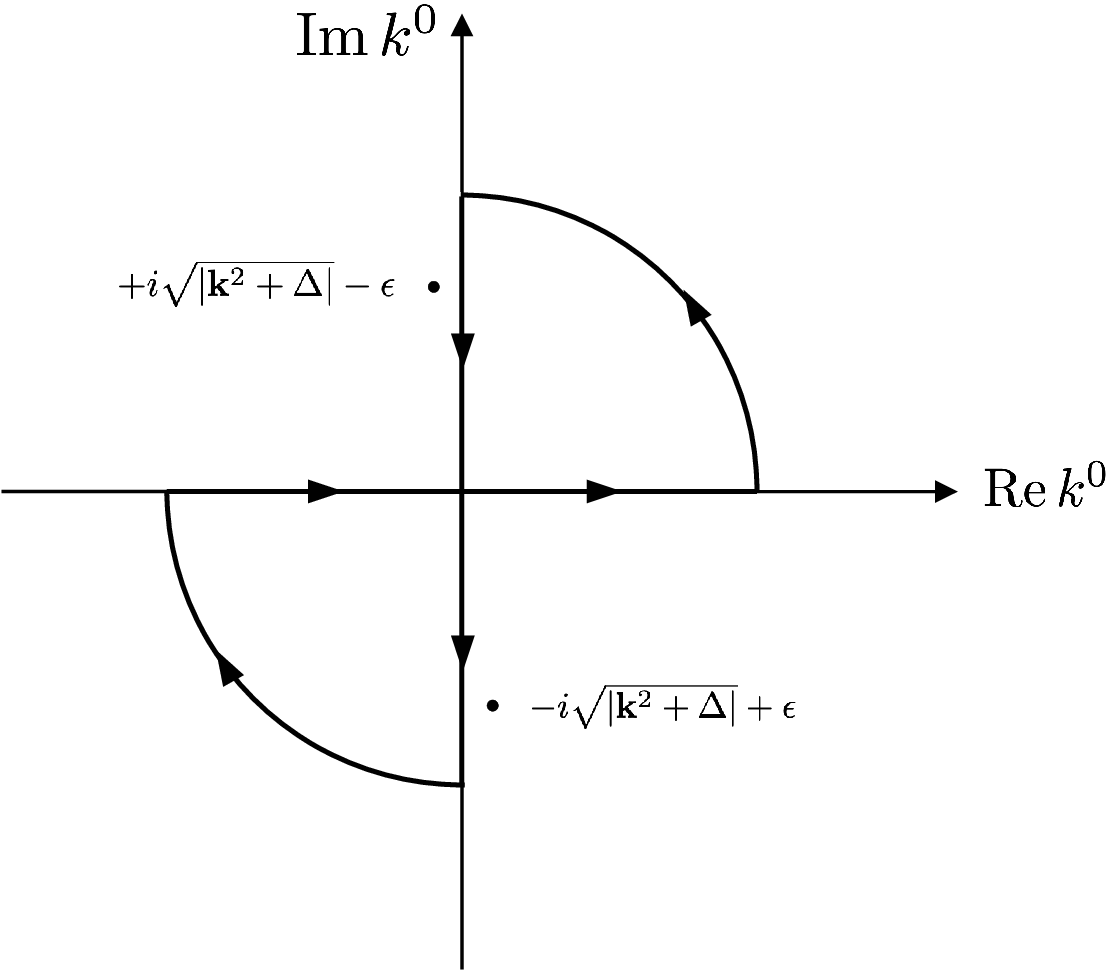
     }
     \caption{}
   \end{subfigure}

\caption{Contour of integration for Wick rotation, relevant to the 
$2 \rightarrow 2$ process of Fig.~\ref{fig:Mom22} for the cases of: (a)~$\mathbf{k}^2 + \Delta > 0$,
and (b)~$\mathbf{k}^2 + \Delta < 0$.}
\label{fig:contour}
\end{figure}

To make further progress, there is an extra idea needed here.
For concreteness, let us consider the case $\mathbf{k}^2 + \Delta > 0$, 
when our integrand has poles at $\pm \sqrt{\mathbf{k}^2 + \Delta} \mp i\epsilon$. 
The issue is that as $\epsilon \rightarrow 0^+$ we get in trouble: the poles approach the contour
of integration (which here is the real line). Dyson's insight was
to work on the complex-$k^0$ plane and 
to replace the contour
of integration by one which does not include the poles even when 
$\epsilon \rightarrow 0^+$: this is precisely what is shown
in the left panel of Fig.~\ref{fig:contour}. 
The contour does not contain any poles, so we can use Cauchy's residue theorem 
(splitting our curve into two simple closed curves).
Thus, this leads to:
\begin{equation}
\int_{-\infty}^{+\infty} dk^0 \frac{1}{\left [ (k^0)^2 - \mathbf{k}^2 - \Delta + i \epsilon \right ]^2} =
\int_{-i\infty}^{+i\infty} dk^0 \frac{1}{\left [ (k^0)^2 - \mathbf{k}^2 - \Delta \right ]^2}
\label{eq:massreno2}
\end{equation}
Observe that on the right-hand side our integral is over the imaginary axis and the 
poles never approach the imaginary axis, so we can safely take 
the $\epsilon \rightarrow 0^+$
limit there.
This counter-clockwise rotation from the real axis to the imaginary axis is now
called a \textit{Wick rotation}. At this point (but not earlier!)
we can introduce the change of variables 
$k^0 \equiv i k_E^0$
in order to go back to the real axis:
\begin{equation}
\int_{-i\infty}^{+i\infty} dk^0 \frac{1}{\left [ (k^0)^2 - \mathbf{k}^2 - \Delta \right ]^2} = 
i\int_{-\infty}^{+\infty} dk_E^0 \frac{1}{\left [ -(k_E^0)^2 - \mathbf{k}^2 - \Delta  \right ]^2}
\label{eq:massreno3}
\end{equation}
We are gratified that we have been able to produce the nicely symmetric
term $(k_E^0)^2 + \mathbf{k}^2$, but also nice (real) integration limits.

We now turn to the second sub-fallacy in our quote: this revolves
around the absence of the $\epsilon$ in Eq.~(\ref{eq:wickrot1}).
It is possible that $\mathbf{k}^2 + \Delta < 0$, in which case 
the poles $\pm i \sqrt{|\mathbf{k}^2 + \Delta|} \mp \epsilon$ are close to the imaginary axis, as illustrated in the right panel of Fig.~\ref{fig:contour}.
This means that, while we are free to carry out the same rotation
of the integration contour, we are not allowed to drop the $\epsilon$ on 
the right-hand side like we did in Eq.~(\ref{eq:massreno2}). 
This issue is related (yet distinct from) another
complication, also quite prevalent in QFT textbooks, namely that when $\Delta < 0$ dropping the $\epsilon$
leads to the logarithm of a negative number (not a pleasant sight). By keeping the 
$\epsilon$, we can treat this as the natural logarithm of a complex number 
and thereby follow the right branch, i.e.,  
$\ln (x - i \epsilon) = \ln|x| - i \pi$ for $x<0$.

{\bf Correction \#6} \textit{Wick rotation is an idea put forward by Freeman Dyson in 1949 to simplify the evaluation of loop integrals, by rotating the contour counter-clockwise. It is not a simple change of variables. You can often drop the infinitesimal in the denominator after you’ve carried out the Wick rotation but, if you always do so, you will get in trouble (e.g., having to take the logarithm of a negative number).}

\section{Summary and conclusion}

In this article, we have discussed in some detail several 
conceptual misunderstandings that arise in introductory textbook treatments
of quantum field theory. They range from themes relevant to classical field
theory (or relativistic QM), to the relationship between classical and 
quantum field theory formulations, all the way to QFT proper. Some of them
can be grasped even by a beginner, while others require a bit of background
in order to be appreciated properly.
A unifying
thread is that these are all topics that a student cannot be reasonably
expected to figure out on their own, especially when the standard modern 
textbook discussions are either silent or erroneous
on the conceptual core of each theme. 

We now tentatively put forward some conjectures on 
why these mistakes arose in the first place.
Some are the result of swiftly dispatching the subject's
preliminaries (Mistake \#1),
while others are the result of sloppiness (Mistake \#6). At least one of them feels
like an individual author's understandable discomfort getting the better of them
(Mistake \#2), with the error later propagating through the literature. Most result
from not sticking to conceptual distinctions consistently (Mistakes \#3 and \#4)
or sometimes not taking the time to wonder why a given distinction should
be made in the first place (Mistake \#5). Of course, the (hypothetical) history of these misconceptions is much less interesting than the task of eliminating them.

A common theme, visible in the references that we cited while
correcting each misconception, is that works that were published several decades ago are typically more careful than the textbooks currently used 
to teach the subject. 
Ours is not a treatise on sociology, but it is not unreasonable to propose that part of the problem comes from a culture valorizing novelty, often at the expense
of depth of understanding. More mundanely, some of the 
pressure working against a profound understanding of the fundamentals of QFT
probably comes from the need to 
cover new material without significantly increasing a textbook's page count.
In the study of neural networks, one encounters
the concept of ``catastrophic forgetting;'' clearly, this is an idea that has 
wider applicability. We hope that the present article, by going over some fairly subtle yet foundational 
issues, will help instructors remember
(or even just learn) correct approaches to 
these themes, thereby indirectly advancing the 
quality of the education available to students of QFT.

\begin{acknowledgments}

This work was supported by the Natural Sciences and 
Engineering Research Council (NSERC) of Canada and the
Canada Foundation for Innovation (CFI).

\end{acknowledgments}


\begin{thebibliography}{99}

\bibitem{GezerlisWilliams1} 
A. Gezerlis and M. Williams, ``Six textbook mistakes in computational physics'', Am. J. Phys. \textbf{89}, 51-60, (2021).

\bibitem{GezerlisWilliams2} 
A. Gezerlis and M. Williams, ``Six textbook mistakes in data analysis'', Eur. Phys. J. Plus \textbf{138}, 19, (2023).

\bibitem{GezerlisNumerical1} 
A. Gezerlis, \textit{Numerical Methods in Physics with Python}, (Cambridge University Press, 2020). 

\bibitem{GezerlisNumerical2} 
A. Gezerlis, \textit{Numerical Methods in Physics with Python}, 2nd ed. (Cambridge University Press, 2023). 

\bibitem{GezerlisQFT} 
A. Gezerlis, \textit{A Gentle Introduction to Quantum Field Theory} (Cambridge University Press, 2026). 

\bibitem{Aitchison} I. J. R. Aitchison and A. J. G. Hey, \textit{Gauge Theories in Particle Physics}, Vol. I (Institute of Physics Publishing, 2003).

\bibitem{Alvarez} L. {\'A}lvarez--Gom{\'e} and M. {\'A}. V{\'a}zquez-Mozo, \textit{An Invitation to Quantum Field Theory} (Springer, 2012).

\bibitem{Banks} T. Banks, \textit{Modern Quantum Field Theory} (Cambridge University Press, 2008).

\bibitem{Baulieu} L. Baulieu, J. Iliopoulos, and R. S{\'e}n{\'e}or,  \textit{From Classical to Quantum Fields} (Oxford University Press, 2017).

\bibitem{Bogoliubov} N. N. Bogoliubov and D. V. Shirkov, \textit{Introduction to the Theory of Quantized Fields} (Interscience Publishers, 1959).

\bibitem{Brown} L. S. Brown, \textit{Quantum Field Theory} (Cambridge University Press, 1992).

\bibitem{Coleman} S. Coleman, \textit{Quantum Field Theory Lectures of Sidney Coleman} (World Scientific, 2019).

\bibitem{Das} A. Das, \textit{Lectures on Quantum Field Theory} (World Scientific, 2008).

\bibitem{Donoghue} J. Donoghue and L. Sorbo, \textit{A Prelude to Quantum Field Theory} (Princeton University Press, 2022).

\bibitem{Folland} G. P. Folland, \textit{Quantum Field Theory: a Tourist Guide for Mathematicians} (American Mathematical Society, 2008).

\bibitem{Fradkin} E. Fradkin, \textit{Quantum Field Theory: an Integrated Approach} (Princeton University Press, 2021).

\bibitem{Gelis} F. Gelis, \textit{Quantum Field Theory: From Basics to Modern Topics} (Cambridge University Press, 2019).

\bibitem{Greiner} W. Greiner and J. Reinhardt, \textit{Field Quantization} (Springer, 1996).

\bibitem{Gross} F. Gross, \textit{Relativistic Quantum Mechanics and Field Theory} (Wiley, 2004).

\bibitem{Hatfield} B. Hatfield, \textit{Quantum Field Theory Of Point Particles And Strings} (CRC Press, 1992).

\bibitem{Itzykson} C. Itzykson and J.-B. Zuber, \textit{Quantum Field Theory} (McGraw-Hill, 1980).

\bibitem{Kleinert} H. Kleinert, \textit{Particles and Quantum Fields} (World Scientific, 2016).

\bibitem{Lancaster} T. Lancaster and S. J. Blundell, \textit{Quantum Field Theory for the Gifted Amateur} (Oxford University Press, 2014).

\bibitem{Maggiore} M. Maggiore, \textit{A Modern Introduction to Quantum Field Theory} (Oxford University Press, 2005).

\bibitem{Mandl} F. Mandl and G. Shaw, \textit{Quantum Field Theory}, 2nd ed. (John Wiley \& Sons, 2010).

\bibitem{Nastase} H. N\u{a}stase, \textit{Introduction to Quantum Field Theory} (Cambridge University Press, 2020).

\bibitem{Padmanabhan} T. Padmanabhan, \textit{Quantum Field Theory} (Springer, 2016).

\bibitem{Peskin} M. E. Peskin and D. V. Schroeder, \textit{An Introduction to Quantum Field Theory} (CRC Press, 2018).

\bibitem{Radovanovic} V. Radovanovi{\'c}, \textit{Problem Book in Quantum Field Theory}, 2nd ed. (Springer, 2008).

\bibitem{Ramond} P. Ramond, \textit{Field Theory: A Modern Primer}, 2nd ed. (Westview Press, 1990).

\bibitem{Ryder} L. H. Ryder, \textit{Quantum Field Theory}, 2nd ed. (Cambridge University Press, 1996).

\bibitem{Schwartz}  M. D. Schwartz, \textit{Quantum Field Theory and the Standard Model} (Cambridge University Press, 2014).

\bibitem{Srednicki} M. Srednicki, \textit{Quantum Field Theory} (Cambridge University Press, 2007).

\bibitem{Stone} M. Stone, \textit{The Physics of Quantum Fields} (Springer, 2000).

\bibitem{Talagrand} M. Talagrand, \textit{What is a Quantum Field Theory?} (Cambridge University Press, 2022).

\bibitem{Williams} A. Williams, \textit{Introduction to Quantum Field Theory} (Cambridge University Press, 2023).

\bibitem{Zee} A. Zee, \textit{Quantum Field Theory in a Nutshell}, 2nd ed. (Princeton University Press, 2010).

\bibitem{Zinn-Justin} J. Zinn-Justin, \textit{Quantum Field Theory and Critical Phenomena}, 5th ed. (Oxford University Press, 2021).


\bibitem{Schweber} S. S. Schweber, \textit{An Introduction to Relativistic Quantum Field Theory} (Harper \& Row, 1961).

\bibitem{Sterman} G. Sterman, \textit{An Introduction to Quantum Field Theory} (Cambridge University Press, 1993).

\bibitem{Salam} P. T. Matthews and A. Salam, ``The Renormalization of Meson Theories'', Rev. Mod. Phys., \textbf{23}, 311 (1951).

\bibitem{Roman} P. Roman, \textit{Introduction to Quantum Field Theory} (John Wiley \& Sons, 1969).

\bibitem{Weinberg} S. Weinberg, \textit{The Quantum Theory of Fields}, Vol. I: Foundations (Cambridge University Press, 1995).

\bibitem{Haag} R. Haag, \textit{Local Quantum Physics}, 2nd ed. (Springer, 1996).

\bibitem{Duncan} A. Duncan, \textit{The Conceptual Framework of Quantum Field Theory} (Oxford University Press, 2012).

\bibitem{Wick} G. C. Wick ``The Evaluation of the Collision Matrix'', Phys. Rev., \textbf{80}, 268 (1950).

\bibitem{Bjorken} J. D. Bjorken and S. D. Drell, \textit{Relativistic Quantum Fields} (McGraw-Hill, 1965).

\bibitem{Dyson} F. J. Dyson, ``The $S$ Matrix in Quantum Electrodynamics'', Phys. Rev., \textbf{75}, 1736 (1949).

\bibitem{Wick2} G. C. Wick ``Properties of Bethe--Salpeter Wave Functions'', Phys. Rev., \textbf{96}, 1124 (1954).

\end{thebibliography}
\end{document}